# High Temperature superconductivity in a hyperbolic geometry of complex matter from nanoscale to mesoscopic scale

G. Campi[1,2] and A. Bianconi[1,2,3]

[1]*Institute of Crystallography, CNR, via Salaria Km 29.300, Monterotondo Roma, I-00015, Italy*
[2]*RICMASS Rome International Center for Materials Science Superstripes, via dei Sabelli 119A, 00185 Roma, Italy*
[3]*INSTM, Italian Interuniversity Consortium on Materials Science and Technology, Rome Udr, Italy*

**Abstract**

While it was known that High Temperature Superconductivity appears in cuprates showing complex multiscale phase separation due to inhomogeneous charge density wave (CDW) order, the spatial distribution of CDW domains remained an open question for a long time, because of the lack of experimental probes able to visualize their spatial distribution at mesoscale, between atomic and macroscopic scale. Recently scanning micro X-ray Diffraction (SµXRD) revealed CDW crystalline electronic puddles with a complex fat-tailed spatial distribution of their size. In this work we have determined and mapped the anisotropy of the Charge Density Waves (CDW) puddles in $HgBa_2CuO_{4+y}$ (Hg1201) single crystal. We discuss the emergence of high temperature superconductivity in the interstitial space with hyperbolic geometry, that opens a new paradigm for quantum coherence at high temperature where negative dielectric function and interference between different pathways can help to raise the critical temperature.

Organization of local lattice fluctuations and charge inhomogeneity determines intrinsic physical properties in new functional materials. This heterogeneity varies from the scale of microns to the atomic level [1-3]. Imaging structural fluctuations and inhomogeneity is an interdisciplinary fundamental issue for understanding function-structure relationship in complex materials and for the design of new functional materials. Nowadays, the improved X-ray optics and imaging techniques make possible to image the bulk structure at nanoscale and mesoscale that is the scale length between the atomic and macroscopic world. In particular, Scanning micro X ray diffraction (SµXRD) using focused beams on the micro-scale is able to probe the k-space order in different spatial locations of the material. This allows to unveil the complex maps showing the multiscale structure in materials. In this way, ultrastructure and phase separation have been visualized in complex and heterogeneous materials for applications in different field from biomedicine [4-7] to material science [8-11].

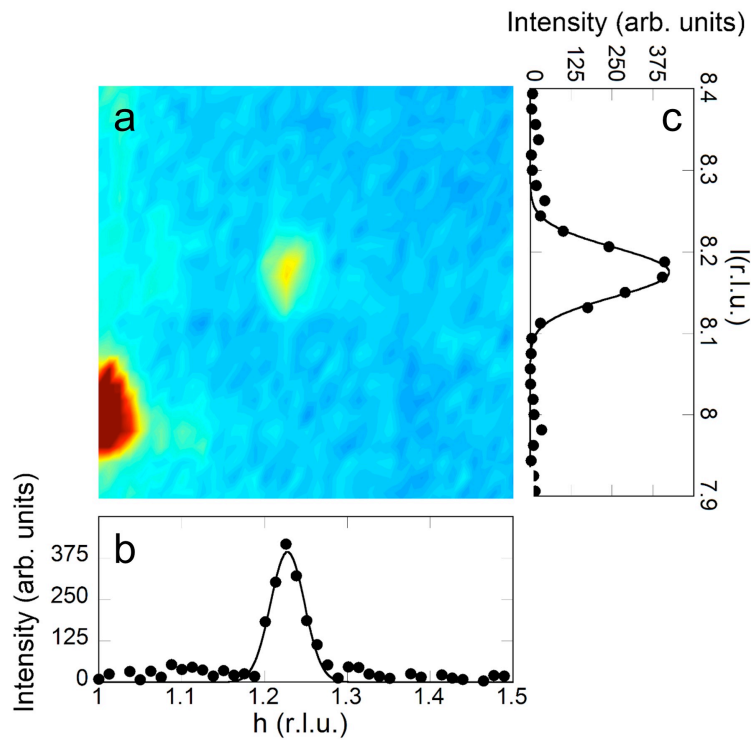

**Figure 1** (a) X ay diffraction pattern measured at ELETTRA showing the CDW peak at q=1.23a* + 8.17c*. X ray profiles of the CDW peak in the (b) h and (c) l directions fitted by Gaussian curves (continuous lines) after background subtraction. The values of FWHM along the h and l direction result to be 0.042(2) and 0.071(1), respectively.

In High Temperature Superconductors (HTS) the spatial imaging of ubiquitous nanoscale phase separation due to quenched disorder of dopants has been obtained both in cuprates [12-14] and iron based superconductors [15-17]. Recently, the interest of scientific community has been focused on the competition of short range charge density wave (CDW) order and superconductivity [18-21]. Scanning micro X-ray diffraction has been used for imaging the spatial distribution of short-range charge-density-wave puddles in $La_2CuO_{4+y}$ [20] and recently in $HgBa_2CuO_{4+y}$ [21] the single-layer cuprate with the highest $T_c$, 95 kelvin [21]. In this work we investigate the anisotropy due to the different size of CDW puddles in the $CuO_2$ plane and out of plane directions.

The Hg1201 crystal was grown at ETH [22]. Crystal structure has P4/mmm symmetry with lattice parameters a=b=0.3875(5) nm and c=0.9508(2) nm at T=100K. We have identified the CDW order by single crystal X-ray diffraction using beam size of 200x200 μm$^2$ and photon energy of 17 KeV on the XRD1 beamline at ELETTRA synchrotron, Trieste [23].

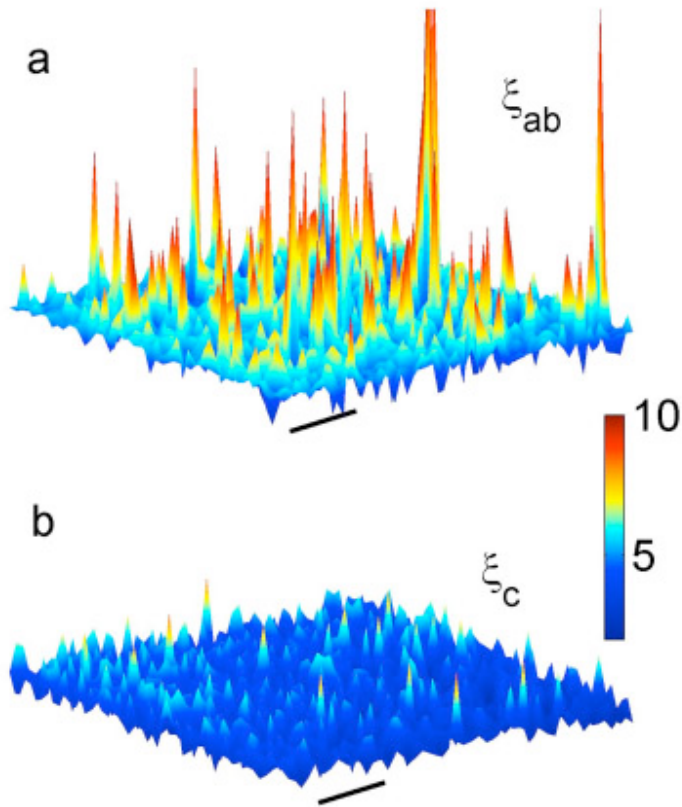

**Figure 2**: Maps of the (a) in plane domain size $\xi_{ab}$ and the (b) out of plane domain size $\xi_c$ The bars correspond to 10 μm.

Only selected reflections show clear CDW satellites, in agreement with Croft [18]. We focused on the CDW satellite $q_{CDW}$=(0.23, 0, 0.17) around the (1 0 8) Bragg reflection, due to stripe-like modulation propagating along the tetragonal a(b)-axis with an average of only 3–4 oscillations per puddle [20]. The CDW order appears below the onset temperature $T_{CDW}$=240 K [20]. **Figure 1a** shows the CDW satellite peak in a diffraction pattern measured at T=100K. By inspection of the profiles of the CDW peak, along the h(k) and l directions, shown in Figure 1b and 1c respectively, we clearly observe how the peak width is larger along the l direction.

Here we determine the variation of CDW puddles size both along the in plane direction and along the out of plane direction, point by point by means of Scanning micro X-ray diffraction (SµXRD) measurements. The sample was kept at T=100K where the system shows the maximum CDW order [20]. SµXRD measurements were performed on the ID13 beamline at ESRF, Grenoble, France.

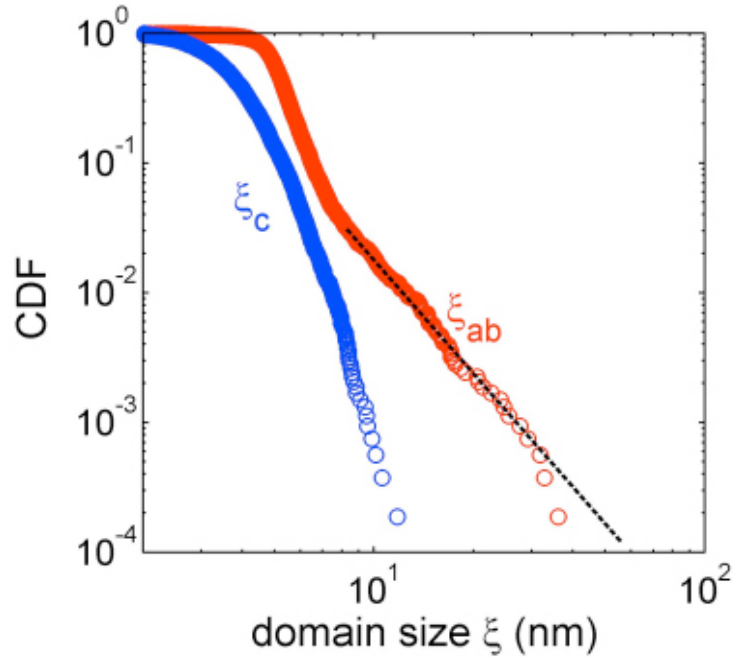

**Figure 3**: Cumulative Density Function of in plane domain size, $\xi_{ab}$ and of the out of plane size $\xi_c$. The in plane domain size shows a fat tail fitted by a power law in the maximum likelihood statistics approach.

We applied an incident X-ray energy of 13 KeV to measure CDW ordering reflections. Moving the sample under a 1x1 µm$^2$ focused beam with an x-y translator, we scanned a sample area of 50x50 µm$^2$ collecting 2500 different diffraction patterns. For each scanned point of the sample the (0.23 0.0 0.16) peak profile has been extracted; the full-width half-maximum (FWHM(a*)) along a* and FWHM(c*) along c* directions have been evaluated to obtain the domain size of the charge ordered puddles both along *a* and *c* crystallographic axis. We got clear evidence of inhomogeneous spatial distribution of

CDW, by mapping the in plane domain size, $\xi_{ab}$=a(b)/FWHM(a*), and the out of plane domain size, $\xi_c$=c/FWHM(c*), as a function of position in the sample.

The map of the CDW domain size, $\xi_{ab}$, and $\xi_c$, are shown in **Figure 2a** and **Figure 2b**, respectively. It is clear that while the puddle size show large fluctuations in the $CuO_2$ planes, the puddle size along the out of plane direction appears uniform. The spatial distributions of puddle size have been characterized by the cumulative probability density function given by $CDF(x) = \int_{-\infty}^{x} p(\xi)d\xi$ where $p(\xi)$ is the probability density function of the size, $\xi$. In **Figure 3** we show the CDF of the domain size, $\xi_{ab}$, and $\xi_c$. We found a fat tailed CDW size distribution along the in plane direction a(b). To investigate whether and how much this upper tail approaches a power-law pattern, we fitted the domain size, $\xi$, with the distribution $p(\xi) = \xi^{\alpha}$ where α is the maximum likelihood estimate of the scaling exponent for $\xi$ larger than a lower bound [24]. The fat tail of the in plane CDW domain size distribution follows a power-law behavior: $p(\xi_{ab}) \approx \xi_{ab}^{-\alpha(CDW)}$ where α(CDW)=2.8±0.2 is the critical exponent. This particular behavior highlights a scale-free like distribution of the in plane CDW domain size in Hg1201 at optimum doping.

From the CDF(x) we can determine that the "average size" of CDW puddles amounts to 4.3 nm in good agreement with previous work probing only the" average size". On the other hand, our space resolved SµXRD measurements indicate a size distribution with fractal-like geometry with rare domains reaching a maximum domain size of 40 nm.

At the same time, the probability distribution of the out of plane CDW size appears to be exponential, indicating a normal size distribution of out of plane CDW puddles. The average out of plane size of the CDW puddles has been found to be about 4 nm.

A view of CDW puddles packing, is represented in **Figure 4a**. The CDW crystalline puddles arrangement introduces a topological change in the available space for the free electrons. These paths go around the CDW puddles in different ways, creating non-trivial interference patterns for electrons and favoring the emergence of the superconducting phase in this complex space.

The free electrons can travel between two different points, taking different paths that cannot be topologically deformed one into the other. This complex space resulting from the measured power law distribution of the in plane puddle size can be mapped to a hyperbolic space [25,26] as it shown in **Figure 4b** where different possible non-equivalent paths taken by electrons at the Fermi level are shown running in the interface space left by the CDW puddles.

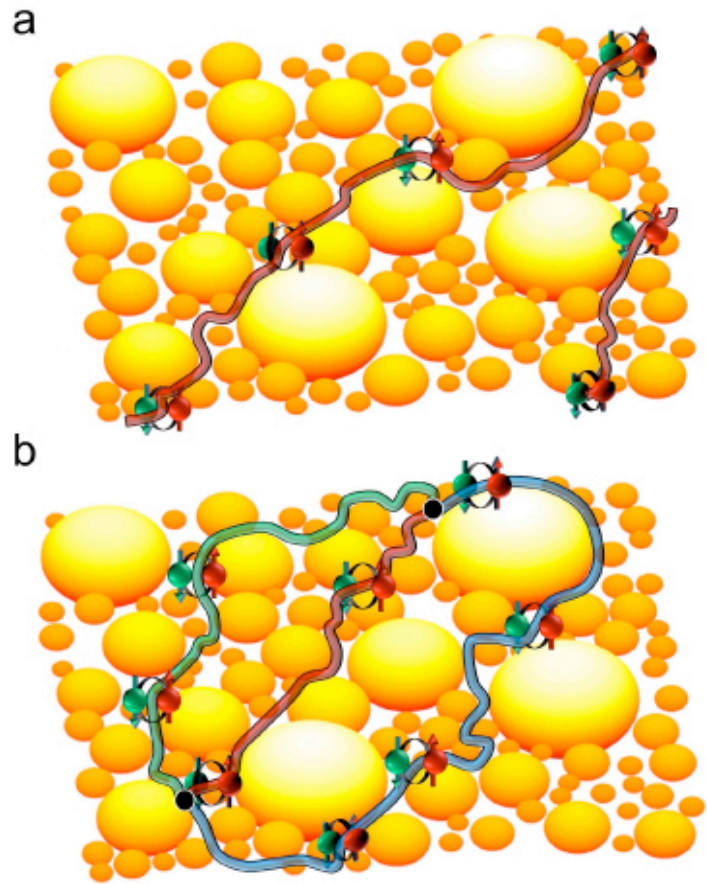

**Figure 4** The CDW crystalline puddles form inhomogeneous spatial patterns giving rise to a new non-Euclidean geometry in the interstitial space left by the crystals of electrons. The free electrons, which do not crystallize, form Cooper pairs flowing along paths in the interstitial space at low temperatures. (a) Pictorial view of the free electrons liquid flowing in the interstitial space left by CDW crystalline puddles. (b) non-equivalence of pathway between two points.

It was already established, using XANES spectroscopy [27-29], that the cuprate superconductors show a heterogeneous copper oxide plane due to polaronic charge density wave formation [30,31]

The use of scanning micro x-ray diffraction [20] has provided compelling evidence for the emergence of the hyperbolic space in the mesoscopic space with hyperbolic geometry favoring superconductivity in cuprates superconductors.

The hyperbolic space has recently been found to play a key role in different fields looking for the emergence of order in complex systems ranging from network theory to quantum gravity [32,33]. The role of a spatial hyperbolic geometry in disordered media is known to be of high relevance in the design of new metamaterials showing negative dielectric constant [34-37], as well as in the manipulation of complex light in optical metamaterials [38-45]. The hyperbolic space has recently been shown to influence critical phenomena [46]. The search for complex materials with zero or negative dielectric constant as been one of the road maps proposed for the amplification of the superconducting critical temperature to reach room temperature superconductivity [47-53]. The present experiments show a hyperbolic space in the mesoscale of cuprates but it could be present also in pressurized sulfur hydride superconductors [54,55], iron based superconductors [56] and diborides [57] which all show local lattice fluctuations. Therefore further investigations are needed to investigate the substantial role of non Euclidean space in the mesoscale range in high temperature superconductors [49-52].